\documentclass[a4paper,10pt,twoside]{cpc-hepnp}

\usepackage{multicol}
\usepackage{graphicx}
\usepackage{amssymb,bm,mathrsfs,bbm,amscd}
\usepackage[tbtags]{amsmath}
\usepackage{lastpage}

\newcommand{\mumu}{\mu^+\mu^-}
\newcommand{\pipi}{\pi^+\pi^-}

\newcommand{\intl}{\int\limits}

\newcommand{\Tau}{$\tau$}
\newcommand{\pp}{$\pi\pi$}

\newcommand{\epem}{$e^+e^-$}
\newcommand{\tautopipiz}{$\tau \rightarrow \nu_\tau \pi \pi^0$}
\newcommand{\amu}{$a_\mu$}
\newcommand{\amuhadLO}{$a_\mu^{\rm had,LO}$}
\newcommand{\amuhadLOpp}{$a_\mu^{\rm had,LO}[\pi\pi]$}
\newcommand{\amuhadLOppzz}{$a_\mu^{\rm had,LO}[\pi\pi2\pi^0]$}
\newcommand{\amuhadHO}{$a_\mu^{\rm had,HO}$}
\newcommand{\tmten}{10^{-10}}
\newcommand{\ea}{{\it et al.}}
\newcommand{\nut}{\nu_\tau}
\newcommand{\beq}{\begin{equation}}
\newcommand{\eeq}{\end{equation}}
\newcommand{\ie}{{\it i.e.}}
\newcommand{\Sew}{S_{EW}}
\newcommand{\Gem}{G_{EM}}

\input babarsym

\begin{document}

\fancyhead[co]{\footnotesize M. Davier: Hadronic Vacuum Polarization and Muon $g-2$}

\footnotetext[0]{Received December 2009}

\title{New Results on the Hadronic Vacuum Polarization Contribution to the Muon $g-2$}

\author{%
      Michel Davier$^{1)}$\email{davier@lal.in2p3.fr}%
}
\maketitle

\address{%
Laboratoire de l'Acc{\'e}l{\'e}rateur Lin{\'e}aire,
             IN2P3/CNRS, Universit\'e Paris-Sud 11, Orsay 91898, France
}

\begin{abstract}
Results on the lowest-order hadronic vacuum polarization contribution to 
the muon magnetic anomaly are presented. They are based on the latest published
experimental data used as input to the dispersion integral. Thus recent results
on \tautopipiz\ decays from Belle and on \epem\ annihilation to $\pi^+ \pi^-$ 
from \babar\ and KLOE are included. The new data, together with improved 
isospin-breaking corrections for $\tau$ decays, result into a much better
consistency among the different results. A discrepancy between the Standard 
Model prediction and the direct $g-2$ measurement is found at the level 
of $3\sigma$.
\end{abstract}


\begin{pacs}
{13.40Em, 13.60.Hb, 13.66.Bc, 13.66.Jn}
\end{pacs}

\begin{multicols}{2}

\section{Introduction}

The Standard Model (SM) prediction of the anomalous magnetic moment of the 
muon, \amu, is limited in precision by contributions from hadronic vacuum 
polarisation (HVP) loops. These contributions can be conveniently separated 
into a dominant lowest order (\amuhadLO) and higher order (\amuhadHO) parts. 
The lowest order term can be calculated with a combination of experimental 
cross section data involving \epem\ annihilation to hadrons, 
and perturbative QCD.
These are used to evaluate an energy-squared dispersion integral, ranging 
from the $\pi^0\gamma$ threshold to infinity. The integration kernel strongly 
emphasises the low-energy part of the spectrum, dominated by the \pp\ final 
state.\footnote{Throughout this paper, final state photon radiation is implied 
for hadronic final states.} When using \epem\ data a deviation of more than 
$3\sigma$ was observed~\cite{md_tau06,hmnt,jegerproc} between the SM prediction
and the direct experimental value~\cite{bnl}.

A former lack of precise \epem-annihilation data inspired the search 
for an alternative. It was found~\cite{adh} in form of 
$\tau\to\nut+{\rm hadrons}$ spectral functions, transferred from the charged 
to the neutral state using isospin symmetry. During the last decade, new 
measurements of the \pp\ spectral function in \epem\ annihilation with percent 
accuracy became available, superseding or 
complementing older and less precise data. With the increasing precision, 
which today is on a level with the $\tau$ data in that channel, 
systematic discrepancies in shape and normalisation of the spectral functions 
were observed between the two systems~\cite{dehz02,dehz03}. It was found that, 
when computing the hadronic VP contribution to the muon magnetic anomaly 
using the \Tau\ instead of the \epem\ data for the $2\pi$ and $4\pi$ channels, 
the observed deviation with the experimental value would 
reduce to less than $1\sigma$~\cite{md_tau06}. Fig.~\ref{amures-06} summarizes
the comparison between theory and experiment by 2006-8~\cite{md_tau06}.

\begin{center}
  \includegraphics[width=7cm]{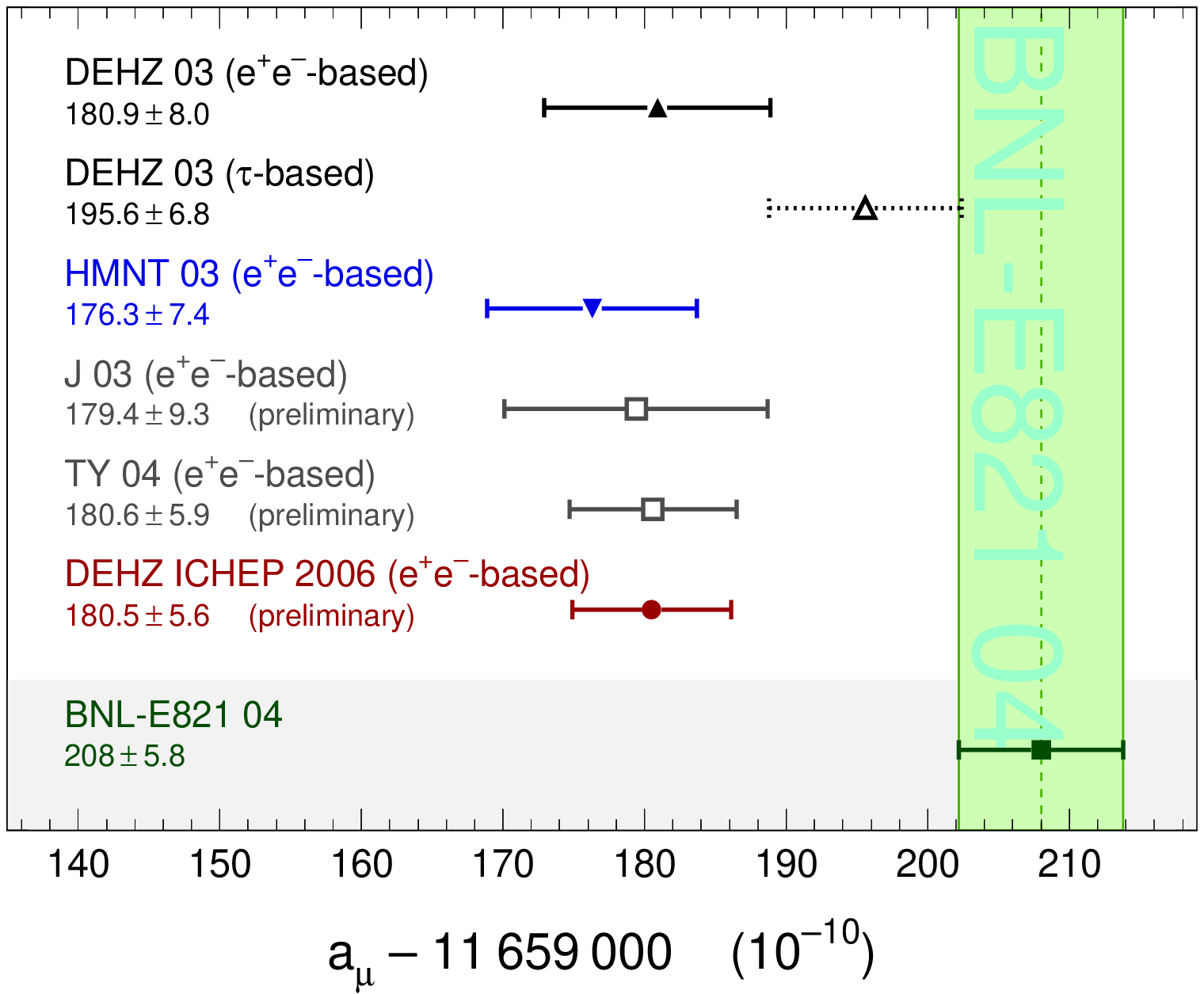}
  \figcaption{\label{amures-06} \small
	Comparison of the predictions for the muon magnetic anomaly with the 
	BNL measurement~\cite{bnl} in 2006, from (top to bottom) 
        Refs.~\cite{dehz03,teubner,yndurain,jegerlehner,md_tau06}.}
\end{center}

In this review I present the situation as of October 2009, taking advantage
of very recent papers: (1) an updated analysis~\cite{new-g-2-tau} using \Tau\ 
data, including high-statistics Belle results~\cite{belle} and an improved
treatment of isospin-breaking corrections (IB)~\cite{lopez-phipsi}; 
(2) a \babar\ measurement~\cite{babarpipi} of the \pp\ spectral function 
using the hard initial state radiation (ISR) method, benefiting from a large 
cancellation of systematic effects in 
the ratio $\pi\pi\gamma(\gamma)$ to $\mu\mu\gamma(\gamma)$ employed for the 
measurement; and (3) a global analysis~\cite{new-g-2-ee} of all published 
\epem data.

\section{HVP and $g-2$}
\label{prediction}

It is convenient to separate the Standard Model (SM) prediction for the
anomalous magnetic moment of the muon
into different contributions,
\begin{equation}
    a_\mu^{\rm SM} \:=\: a_\mu^{\rm QED} + a_\mu^{\rm had} +
                             a_\mu^{\rm weak}~,
\end{equation}
with
\begin{equation}
 a_\mu^{\rm had} \:=\: a_\mu^{\rm had,LO} + a_\mu^{\rm had,HO}
           + a_\mu^{\rm had,LBL}~,
\end{equation}
and where $a_\mu^{\rm QED}=(11\,658\,471.810\pm0.016)~\tmten$ is 
the pure electromagnetic contribution~\cite{kinoshita}, 
\amuhadLO\ is the lowest-order HVP contribution, 
$a_\mu^{\rm had,HO}=(-9.79\pm0.08_{\rm exp}\pm0.03_{\rm rad})~\tmten$ 
is the corresponding higher-order part~\cite{krause2,teubner}, 
and $a_\mu^{\rm weak}=(15.4\pm0.1\pm0.2)~\tmten$,
where the first error is the hadronic uncertainty and the second
is due to the Higgs mass range, accounts for corrections due to
exchange of the weakly interacting bosons up to two loops~\cite{amuweak}. 
For the light-by-light (LBL) scattering part, $a_\mu^{\rm had,LBL}$,
we use the value $(10.5\pm2.6)~\tmten$ from the latest 
evaluation~\cite{prades09}.

Owing to unitarity and to the analyticity of the vacuum-polarization 
function, the lowest order HVP contribution to $a_\mu$ can be computed 
through the dispersion integral~\cite{rafael}
\begin{equation}
\label{eq_int_amu}
    a_\mu^{\rm had,LO} \:=\: 
           \frac{\alpha^2}{3\pi^2}
           \intl_{4m_\pi^2}^\infty\!\!ds\,\frac{K(s)}{s}R^{(0)}(s)~,
\end{equation}
where $K(s)$ is a well-known QED kernel, $\alpha=\alpha(s=0)$, and
$R^{(0)}(s)$ denotes the ratio of the ``bare'' cross
section for $e^+e^-$ annihilation into hadrons to the pointlike muon-pair cross
section. The bare cross section is defined as the measured cross section
corrected for initial-state radiation, electron-vertex loop contributions
and vacuum-polarization effects in the photon propagator. However, photon 
radiation in the final state is included in the bare cross section 
defined here. The reason for using the bare (\ie, lowest order) 
cross section is that a full treatment of higher orders is anyhow 
needed at the level of $a_\mu$, so that the use of the ``dressed''
cross section would entail the risk of double-counting some of the 
higher-order contributions.

The function $K(s)\sim1/s$ in Eq.~(\ref{eq_int_amu}) gives a strong 
weight to the low-energy part of the integral. About 91\% of the 
total contribution to \amuhadLO\  is accumulated at centre-of-mass
energies $\sqrt{s}$ below 1.8~GeV and 73\% is covered 
by the $\pi\pi$ final state, which is dominated by the $\rho(770)$ 
resonance. 

\section{Updated $2\pi$ Analysis Using $\tau$ Data}

\subsection{Spectral functions in $\tau$ decays}

The spectral function of 
the vector current decay $\tau\to X^-\nut$ is related to the $\ee\to X^0$ 
cross section of the corresponding isovector final state $X^0$,
\begin{equation}
\label{eq:cvc}
  \sigma_{X^0}^{I=1}(s) =
         \frac{4\pi\alpha^2}{s}\,v_{1,\,X^-}(s)~,
\end{equation}
where $s$ is the centre-of-mass energy-squared or equivalently 
the invariant mass-squared of the $\tau$ final state $X$, 
$\alpha$ is the electromagnetic fine structure constant, 
and $v_{1,\,X^-}$ is the non-strange, isospin-one vector spectral function
corrected for IB and given by 
\begin{eqnarray}
\label{eq:sf}
   v_{1,\,X^-}(s)
   &=&
           \frac{m_\tau^2}{6\,|V_{ud}|^2}\,
              \frac{\BR_{X^-}}
                   {\BR_{e}}\,
              \frac{1}{N_X}\frac{d N_{X}}{ds}  \\
   & & 
              \times\,
              \left(1-\frac{s}{m_\tau^2}\right)^{\!\!-2}\!
                     \left(1+\frac{2s}{m_\tau^2}\right)^{\!\!-1}
              \frac{R_{\rm IB}(s)}{\Sew} \,, \nonumber
\end{eqnarray}
with
\begin{equation}
\label{eq:rib}
R_{\rm IB}(s)=\frac{{\rm FSR}(s)}{\Gem(s)}
              \frac{\beta^3_0(s)}{\beta^3_-(s)}
              \left|\frac{F_0(s)}{F_-(s)}\right|^2\,.
\end{equation}
In Eq.~(\ref{eq:sf}), $(1/N_X)dN_X/ds$ is the normalised invariant mass 
spectrum of the hadronic final state, and $\BR_{X^-}$ denotes 
the branching fraction of $\tau\rightarrow \nut X^-$. 
We use for the $\tau$ mass the value
$m_\tau=(1776.84\pm 0.17)\,{\rm MeV}$~\cite{pdg08}, and for the CKM matrix 
element $|V_{ud}|=0.97418\pm0.00019$~\cite{ckmfitter}, 
which assumes CKM unitarity. For the electron branching fraction we use 
$\BR_{e}=(17.818 \pm 0.032)\%$, obtained~\cite{rmp} supposing lepton 
universality. Short-distance electroweak radiative effects lead to the 
correction $\Sew=1.0235\pm0.0003$~\cite{marciano,dehz02,bl90,erler04}.
All the $s$-dependent IB corrections are included in 
$R_{\rm IB}$, which have been worked out in the $2\pi$ channel: ${\rm FSR}(s)$ 
refers to the final state radiative corrections~\cite{fsr}, and $\Gem(s)$ 
denotes the long-distance radiative corrections of order $\alpha$ to the 
photon inclusive $\tau$ spectrum, computing the virtual and real photonic 
corrections using chiral resonance~\cite{cirigliano} or vector 
dominance~\cite{lopez}.

\subsection{Improved IB corrections}

As the physics of IB is described elsewhere~\cite{lopez-phipsi} I only
concentrate here on the major difference between the previous 
analysis~\cite{dehz02,dehz03} and the new one~\cite{new-g-2-tau}. 
In Eq.~(\ref{eq:rib}) the ratio of the electromagnetic $F_0(s)$ and weak
$F_-(s)$ form factors depends on the charged and neutral $\rho$ parameters, 
as well as $\rho-\omega$ interference. While a small mass difference (of the 
order of 1 MeV) makes only a small effect on the dispersion integral because 
of its bipolar nature, a difference in the width can lead to a significant
correction. Such a difference is expected from radiative 
$\rho\rightarrow\pi\pi\gamma$ decays which have been evaluated in a scalar-QED
vector-dominance model~\cite{lopez-rhoppg}. The result is markedly different
from the estimate made previously using only the hard radiation 
part~\cite{singer}.

One could question the validity of using point-like pions in the calculation
of radiative decays. However several experimental tests support this
assumption in $e^+e^-\rightarrow\pi^+\pi^-\gamma(\gamma)$ for the same mass
range: lowest-order FSR with KLOE~\cite{mueller-phipsi}, additional FSR with 
\babar~\cite{babarpipi}.

The new IB corrections are listed in Table~\ref{tab:amu}.

\subsection{Consistency of $\tau$ spectral functions}

\begin{figure}[htb]
\begin{center}
\includegraphics[width=70mm]{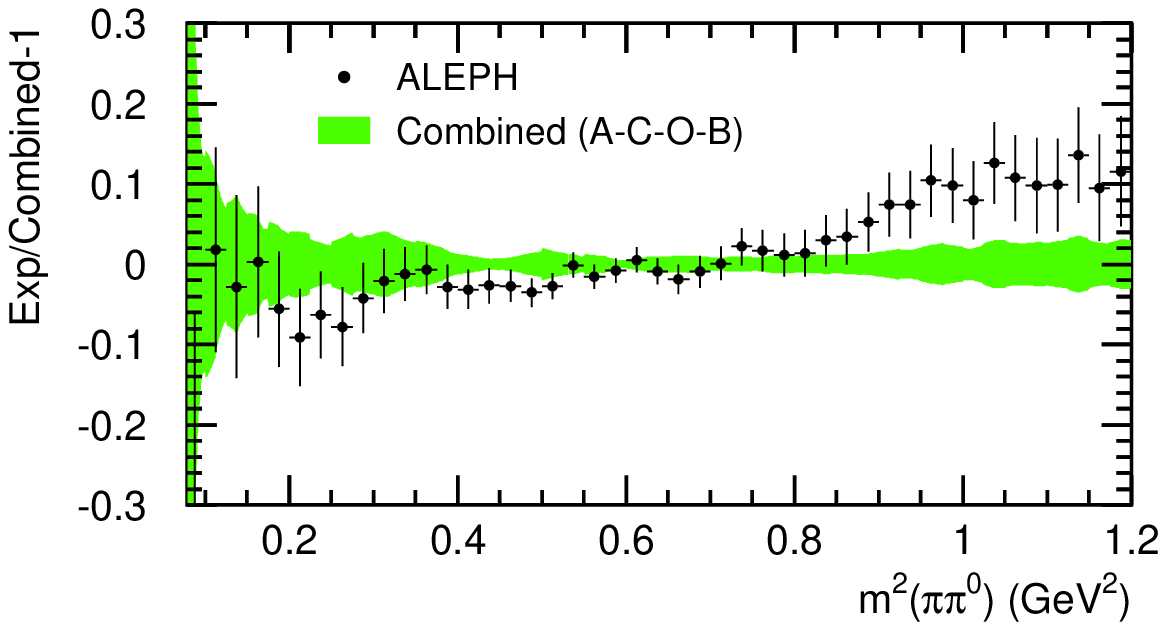}
\includegraphics[width=70mm]{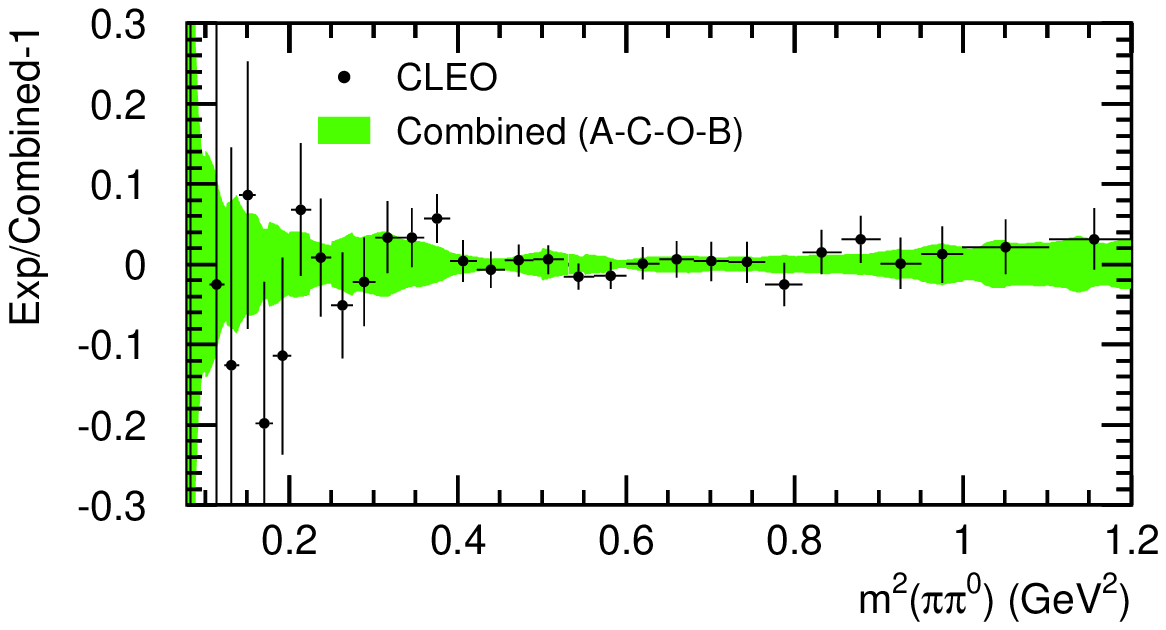} \\
\includegraphics[width=70mm]{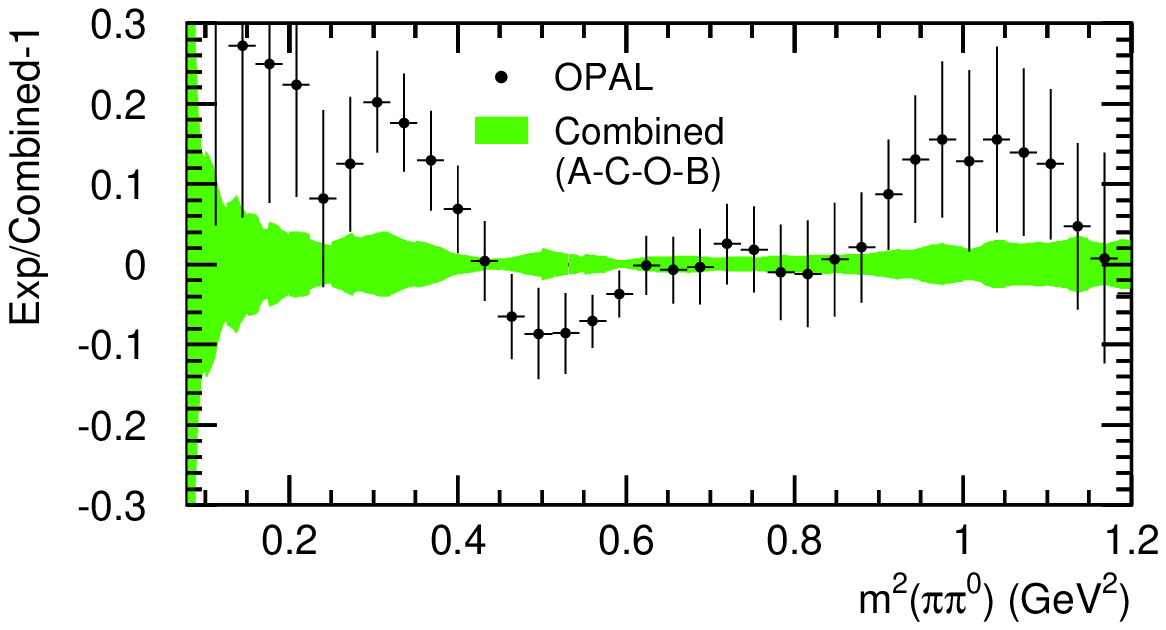} 
\includegraphics[width=70mm]{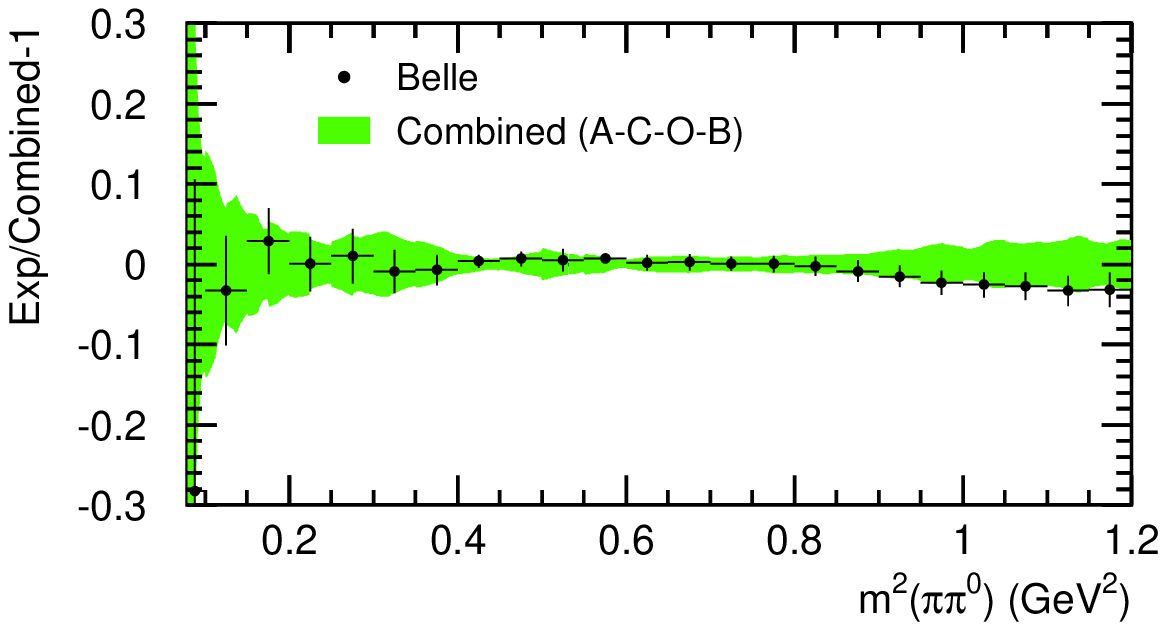}
\end{center}
\vspace{-0.3cm}
\caption{Relative comparison between the $\tau\rightarrow\nut\pi\pi^0$ spectral
functions from ALEPH, CLEO, OPAL, Belle (data points) and 
their combined result (shaded band).}
\label{fig:tau_com}
\end{figure}

All published $\tau$ $2\pi$ spectral functions are normalised to the 
world-average branching ratio. The shape of their mass dependence can be
compared by looking at the relative difference between each spectral function
and the combined one, locally averaging the data from ALEPH~\cite{aleph-2pi},
CLEO~\cite{cleo-2pi}, OPAL~\cite{opal-2pi}, and Belle~\cite{belle} 
(Fig.~\ref{fig:tau_com}).

\begin{center}
\tabcaption{ \label{tab:amu} \small
    Contributions to \amuhadLO\ $[\pi\pi, \tau]$
    from the isospin-breaking corrections.
    Corrections shown in two separate columns correspond to the 
    Gounaris-Sakurai (GS) and K\"uhn-Santamaria (KS) form factor 
    parametrisations, respectively.} 
\small
\begin{tabular}{ccc} \hline\hline
Source      & \multicolumn{2}{c}{$\Delta$ \amuhadLO\ $[\pi\pi, \tau]$ ($10^{-10}$)}\\
            &  GS Model & KS Model  \\ \hline
 $S_{EW}$            & \multicolumn{2}{c}{$-12.21\pm0.15$}\\
 $G_{EM}$            & \multicolumn{2}{c}{$-1.92\pm0.90$}\\
 FSR             & \multicolumn{2}{c}{$+4.67\pm0.47$}\\
$\rho-\omega$ interference & $+2.80\pm0.19$ & $+2.80\pm0.15$ \\
$m_{\pi^\pm}-m_{\pi^0}$ ($\sigma$)  & \multicolumn{2}{c}{$-7.88$}\\
$m_{\pi^\pm}-m_{\pi^0}$ ($\Gamma_\rho$) & $+4.09$ & $+4.02$\\
$m_{\rho^\pm}-m_{\rho^0_{\rm bare}}$ & $+0.20^{+0.27}_{-0.19}$ & $+0.11^{+0.19}_{-0.11}$ \\
$\pi\pi\gamma$, EM decays & $-5.91\pm0.59$ & $-6.39\pm0.64$ \\ \hline
 Total         & $-16.07\pm0.59$ & $-16.70\pm0.64$ \\
               & \multicolumn{2}{c}{$-16.07\pm1.85$}\\
\hline\hline
\end{tabular}
\end{center}

\end{multicols}

\begin{multicols}{2}

Since the world-average branching ratio is dominated by the ALEPH result, 
it is interesting to test the consistency between the absolute spectra, 
{\it i.e.} when each spectrum is normalised to the branching ratio 
measured by the same experiment. 
Fig.~\ref{fig:tau_amu} shows a very good agreement between the full dispersion
integrals with comparable uncertainties. Thus the $\tau$ experiments yield
consistent absolute results. The average value and its error are rather 
insensitive to the branching ratio choice, although it is not true at the level
of individual experiments. In particular the Belle result reaches its best 
precision only when the world-average is used.

\begin{center}
  \includegraphics[width=8cm]{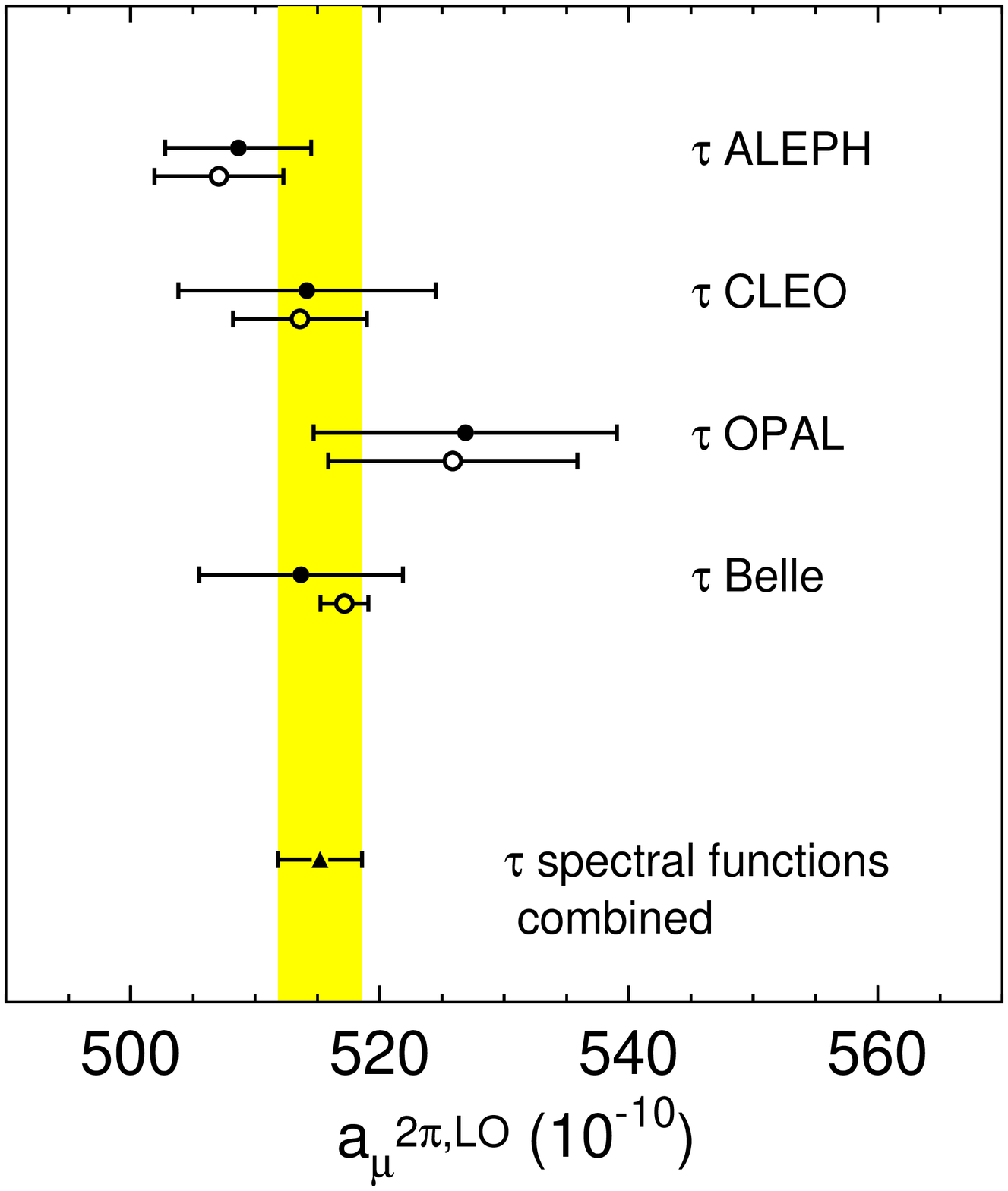}
  \figcaption{\label{fig:tau_amu} \small
	Comparison of the \amuhadLOpp\ values for different $\tau$
experiments using their own $\tau\rightarrow\nut\pi\pi^0$ branching ratios
(closed circles) or the world-average (open circles).}
\end{center}

\subsection{Comparison to $e^+e^-$ Data}

Figure~\ref{fig:tauee} shows the relative difference between the $ee$ 
and the IB-corrected $\tau$ spectral functions versus $s$. 
The relative normalisation is consistent within the respective errors
and the shape is found in better agreement than before~\cite{dehz03}, despite 
a remaining deviation above the $\rho$-mass-squared. The discrepancy with 
the KLOE data, although reduced, persists.
\begin{center}
\includegraphics[width=70mm]{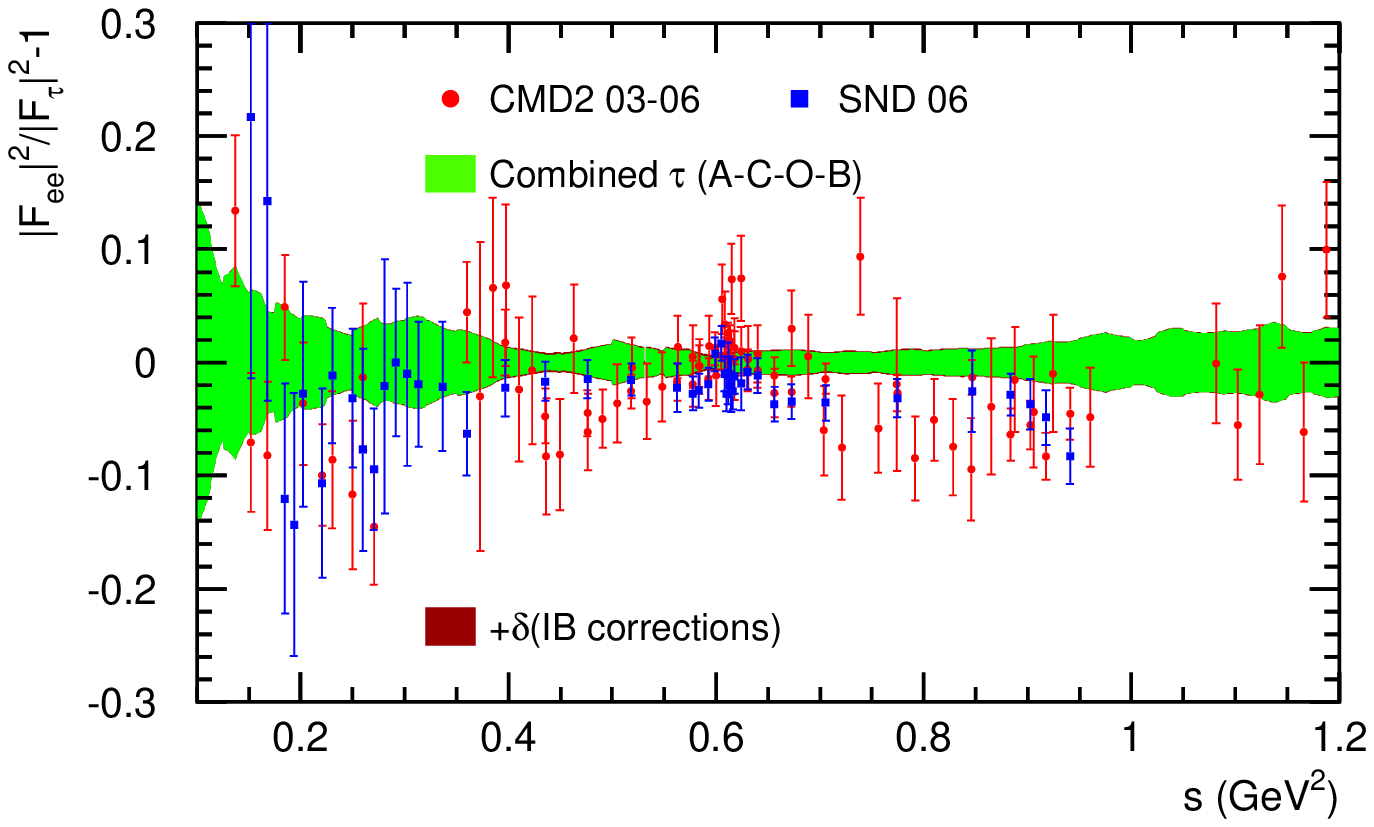}\\
\includegraphics[width=70mm]{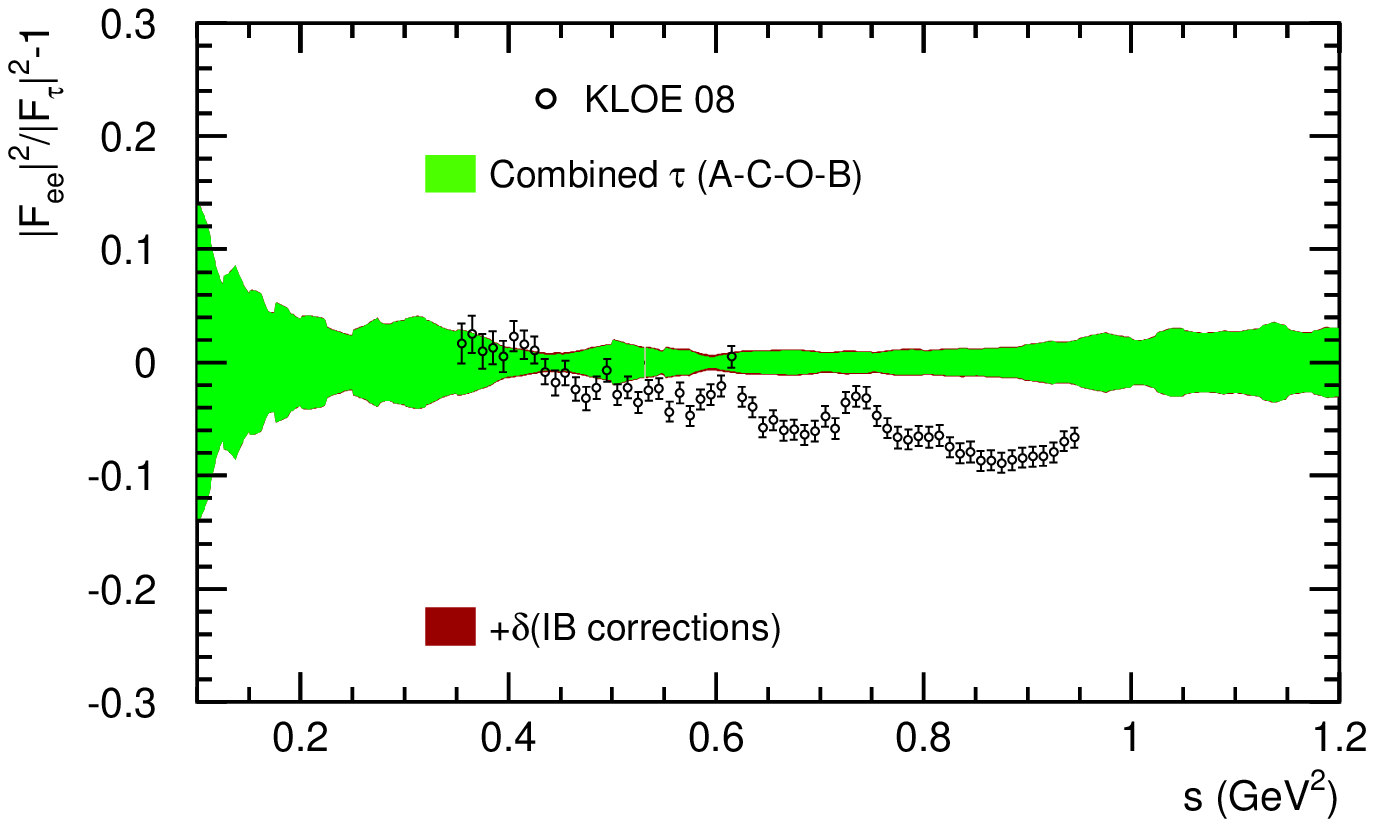}
\figcaption{\label{fig:tauee} \small
Relative comparison between $ee$ and $\tau$ spectral functions, 
   expressed in terms of the difference between neutral and charged pion form 
   factors. Isospin-breaking corrections are applied to $\tau$ data with 
   the corresponding uncertainties included in the error band.}

\end{center}

As the g-2 dispersion relation involves an integral over the hadronic spectral
function, it is interesting to consider the result with another kernel.
By integrating the $e^+e^-$ data weighted by the $\tau$ matrix element 
$C_\tau$, and 
correcting for IB, one obtains the branching ratio $B^{CVC}_{\pi\pi^0}$ 
which can be directly compared to the measurements. Indeed,

\begin{eqnarray}
   B^{\rm CVC}_{\pi\pi^0} &=&  \frac{3}{2}
                          \frac{S_{EW}B_e|V_{ud}|^2}{\pi \alpha^2 m^2_\tau}
                          \int^{m^2_\tau}_{4m_\pi^2}\!\!\! ds\,s\, 
                          \frac {\sigma_{\pi^+\pi^-}}{R_{IB}} C_\tau \\
      C_\tau &=&          \left(1-\frac{s}{m^2_\tau}\right)^{\!\!2}
                          \left(1+\frac{2s}{m^2_\tau}\right)\,.
\end{eqnarray}

The results for $B^{CVC}_{\pi\pi^0}$ from $e^+e^-$ experiments are compared to
the direct measurements in Fig.~\ref{fig:brpipi0}
The average, $(24.78 \pm 0.17_{\rm exp} \pm 0.22_{\rm IB})\%$, differ
from the average $\tau$ branching ratio, $(25.42\pm0.10)$\%, by
$(0.64\pm 0.10_\tau\pm 0.17_{ee}\pm 0.22_{IB})\%$ to be compared to an applied 
IB correction of ($+0.69$)\%. The discrepancy of about $2\sigma$ is 
significantly reduced from the previous analysis~\cite{md_tau06} ($4.5\sigma$).
It should be emphasized that the observed increased deviation above the
$\rho$ mass between $ee$ and $\tau$ spectral functions, essentially driven by
the KLOE data, plays a more significant role in the $B^{CVC}_{\pi\pi^0}$ 
integral, rather than in the $a_\mu$ integral with its much steeper kernel. 
So we expect a better consistency for $g-2$.

\begin{center}
\includegraphics[width=70mm]{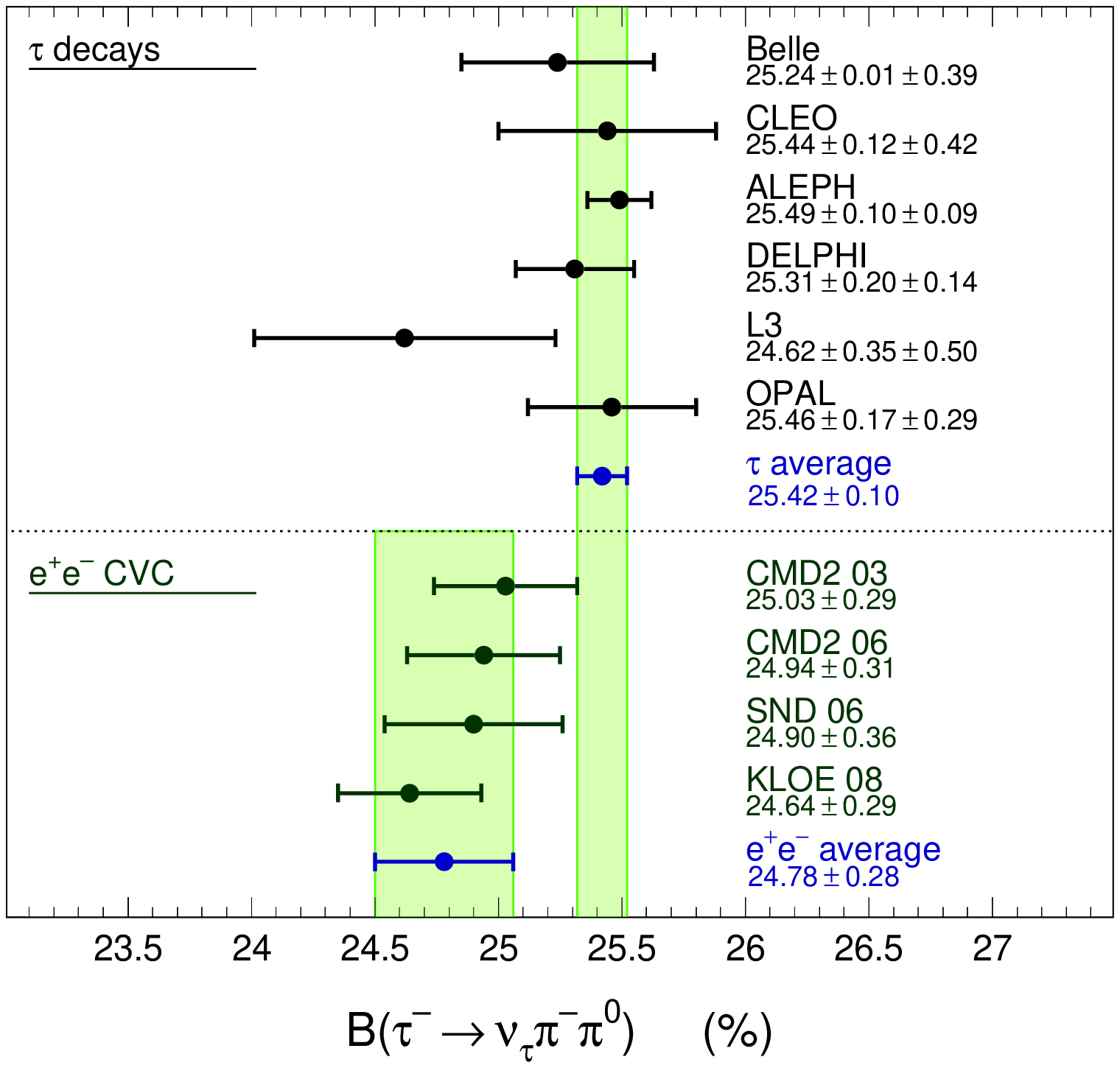}
\figcaption{\label{fig:brpipi0} \small
The measured branching fractions for 
        $\tau\rightarrow\nut \pi\pi^0$ (references in \cite{new-g-2-tau})
        compared to the predictions from the $e^+e^-\rightarrow\pi^+\pi^-$ 
        spectral functions, applying the IB corrections. 
        For the $ee$ results, only the data from the indicated experiments 
        in the $0.63-0.958$~GeV range are used, and 
     the combined $ee$ data elsewhere. Vertical bands indicate average values 
}
\end{center}

\section{Updated $2\pi$  Analysis Using $e^+e^-$ Including \babar}

\subsection{The data}

Recent precision data, where all required radiative corrections have been 
applied by the experiments, stem from the CMD-2~\cite{cmd2new} and 
SND~\cite{snd} experiments at the VEPP-2M collider. They achieve comparable 
statistical errors, and energy-dependent systematic uncertainties down to 
$0.8\%$ and $1.3\%$, respectively.

These measurements have been complemented by results from KLOE~\cite{kloe08} 
at DA$\Phi$NE running at the $\phi$ resonance centre-of-mass energy.
KLOE applied for the first time the ISR technique to precisely 
determine the \pp\ cross section between $0.592$ and $0.975\gev$.  
The high statistics of the analysed data 
sample yields a $0.2\%$ relative statistical error on the \pp\ contribution to 
\amuhadLO. KLOE normalises the $\pi\pi\gamma$ cross section taking the 
absolute ISR radiator function from Monte Carlo simulation 
(Ref.~\cite{phokhara} and references therein). 
The  systematic error  assigned to this correction 
varies between $0.5\%$ and $0.9\%$ (closer to the $\phi$ peak). The total 
assigned systematic error lies between $0.8\%$ and $1.2\%$.

In a recent publication~\cite{babarpipi} the \babar\ Collaboration reported 
measurements of the processes $ee\rightarrow\pi\pi\gamma, \mu\mu\gamma$ using 
the ISR method at 10.6\gev centre-of-mass energy. The detection of the hard 
ISR photon allows \babar\ to cover a large energy range from threshold up to 
$3\gev$ for the two processes. The $\pi\pi(\gamma)$ cross section is obtained 
from the $\pi\pi\gamma(\gamma)$ to $\mu\mu\gamma(\gamma)$ ratio, so that the 
ISR radiation function cancels, as well as additional ISR radiative effects. 
Since additional FSR photons are also detected, there is no additional 
uncertainty from radiative corrections at NLO level. Experimental 
systematic uncertainties are kept to 0.5\% in the $\rho$ peak region 
(0.6--0.9\gev), increasing to 1\% outside. 

\subsection{Combining cross section data}
\label{sec:hvptools}

The details of the combination procedure are given in Ref.~\cite{new-g-2-ee}.
The requirements for averaging and integrating cross section data are: 
($i$) properly propagate all the uncertainties in the data to the final 
integral error, ($ii$) minimise biases, \ie, reproduce the true integral as 
closely as possible in average and measure the remaining systematic error, and 
($iii$) minimise the integral error after averaging while respecting the two 
previous requirements. The first item practically requires the use of 
pseudo-Monte Carlo (MC) simulation, which needs to be a faithful 
representation of the measurement ensemble and to contain the full data 
treatment chain (interpolation, averaging, integration). The second item 
requires a flexible data interpolation method and a realistic truth model used 
to test the accuracy of the integral computation with pseudo-MC experiments. 
Finally, the third item requires optimal data averaging taking into account 
all known correlations to minimise the spread in the integral measured from 
the pseudo-MC sample.

The combination and integration of the $ee\rightarrow\pi\pi$ cross section data
is performed using the newly developed software package 
HVPTools~\cite{hvptools}.
It transforms the bare cross section data and associated statistical and 
systematic covariance matrices into fine-grained energy bins, taking into 
account to our best knowledge the correlations within each experiment as well 
as between the experiments (such as uncertainties in radiative corrections). 
The covariance matrices are obtained by assuming common systematic error 
sources to be fully correlated. To these matrices are added statistical 
covariances, present for example in binned measurements as provided by KLOE, 
\babar\ or the $\tau$ data, which are subject to bin-to-bin migration that has 
been unfolded by the experiments, thus introducing correlations. 
The interpolation between adjacent measurements of a given experiment uses 
second-order polynomials, which is an improvement with respect to the 
previously applied trapezoidal rule. In the case of binned data, the 
interpolation function within a bin is renormalised to keep the integral in 
that bin invariant after the interpolation. The final interpolation function 
per experiment within its applicable energy domain is discretised into 
small (1 MeV) bins for the purpose of averaging and numerical integration. 

The averaging of the interpolated measurements from different experiments 
contributing to a given energy bin is the most delicate step in the analysis 
chain. Correlations between measurements and experiments must be taken into 
account. Moreover, the experiments have different measurement densities or 
bin widths within a given energy interval and one must avoid that missing 
information in case of a lower measurement density is substituted by 
extrapolated information from the polynomial interpolation.
To derive proper weights given to each experiment, wider averaging regions
are defined to ensure that all locally available experiments contribute to the 
averaging region, and that in case of binned measurements at least one 
full bin is contained in it. The averaging regions are used to compute weights 
for each experiment, which are applied in the bin-wise average of the original
finely binned interpolation functions.
If the $\chi^2$ value exceeds the number of degrees of freedom ($n_{\rm dof}$),
the error in the averaged bin is rescaled by $\sqrt{\chi^2/n_{\rm dof}}$ 
to account for inconsistencies. Fig.~\ref{fig:chi2} shows the distributions in
$\sqrt{s}$ of the error recaling factor, the relative weights for each 
experiment, and the contribution to the dispersion integral, as well as its
error. It is seen that \babar\ dominates the averaging up to the $\rho$ peak
and above 0.95 GeV, while KLOE has a larger weight in-between owing to the
steep behaviour of the radiator function when approaching 1 GeV.
The uncertainty in the integral is dominated by the measurements below 0.8\gev.

The consistent propagation of all errors into the evaluation of \amuhadLO\ is 
ensured by generating large samples of pseudo experiments, representing the 
full list of available measurements and taking into account all known 
correlations. For each generated set of pseudo measurements, the identical 
interpolation and averaging treatment leading to the computation of 
Eq.~(\ref{eq_int_amu}) as for real data is performed, hence resulting in a 
probability density distribution for \amuhadLOpp, the mean and RMS of which 
define the $1\sigma$ allowed interval.

The fidelity of the full analysis chain (polynomial interpolation, averaging, 
integration) has been tested with toy models, using as truth representation a 
Gounaris-Sakurai vector-meson resonance model faithfully describing the \pp\ 
data. Negligible biases below 0.1 ($10^{-10}$ units) are found, increasing to 
0.5 (1.2 without the high-density \babar\ data) when using the trapezoidal 
rule for interpolation instead of second order polynomials.

\begin{center}
\includegraphics[width=60mm]{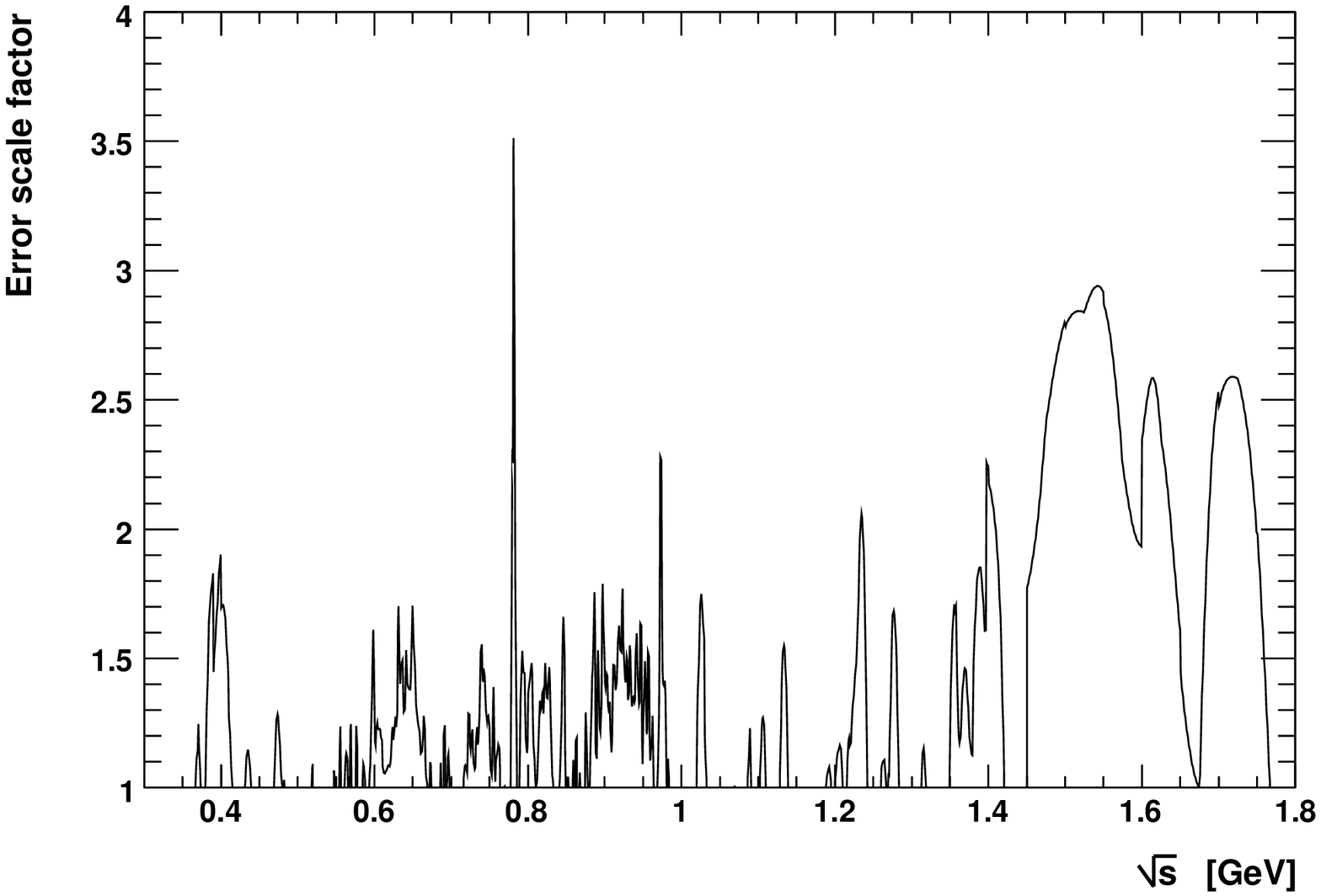}
\includegraphics[width=60mm]{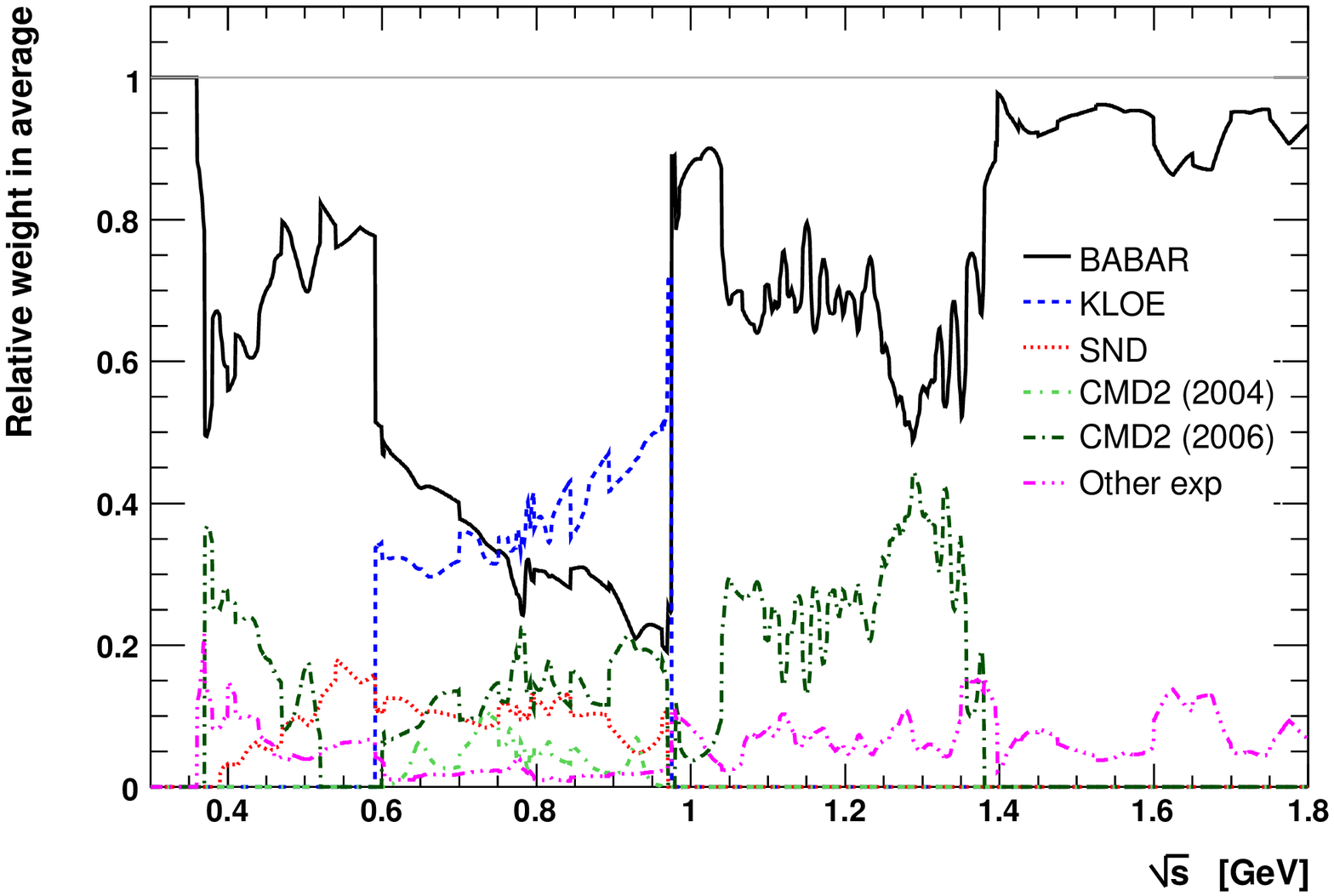}
\includegraphics[width=60mm]{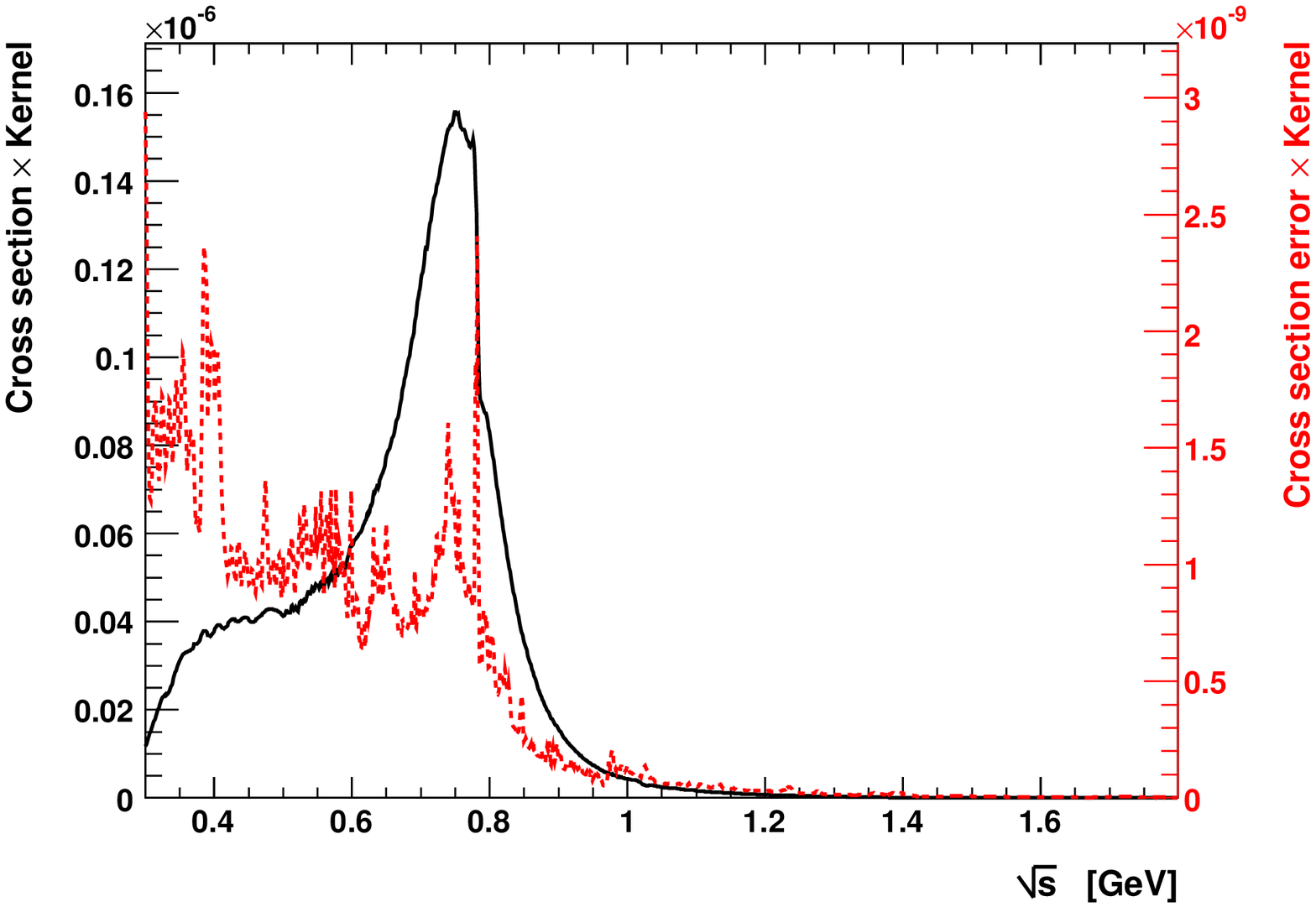}
\figcaption{\label{fig:chi2} \small
           Top: Error rescaling factor accounting for inconsistencies among 
           experiments versus $\sqrt{s}$.
           Middle: Relative averaging weights for each experiment versus 
           $\sqrt{s}$.
           Bottom:  Contribution to the dispersion integral for the 
            combined $ee$ data obtained by multiplying the \pp\ cross section 
           by the kernel function $K(s)$ (solid line). The dashed (red) curve 
            belonging
            to the right axis shows the corresponding error contribution, where
            statistical and systematic errors have been added in quadrature.
}
\end{center}

\end{multicols}

\begin{center}
\includegraphics[width=60mm]{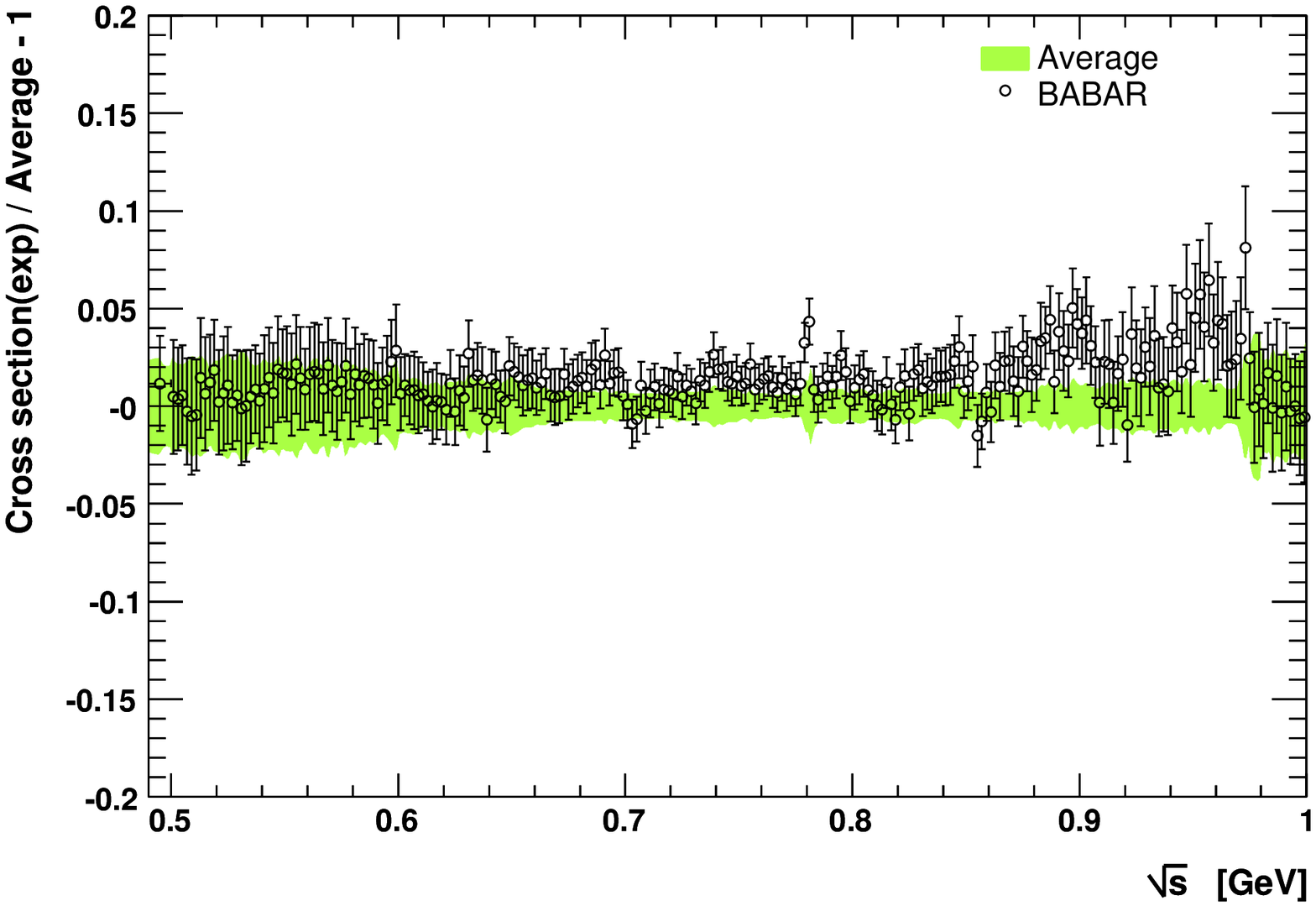}
\includegraphics[width=60mm]{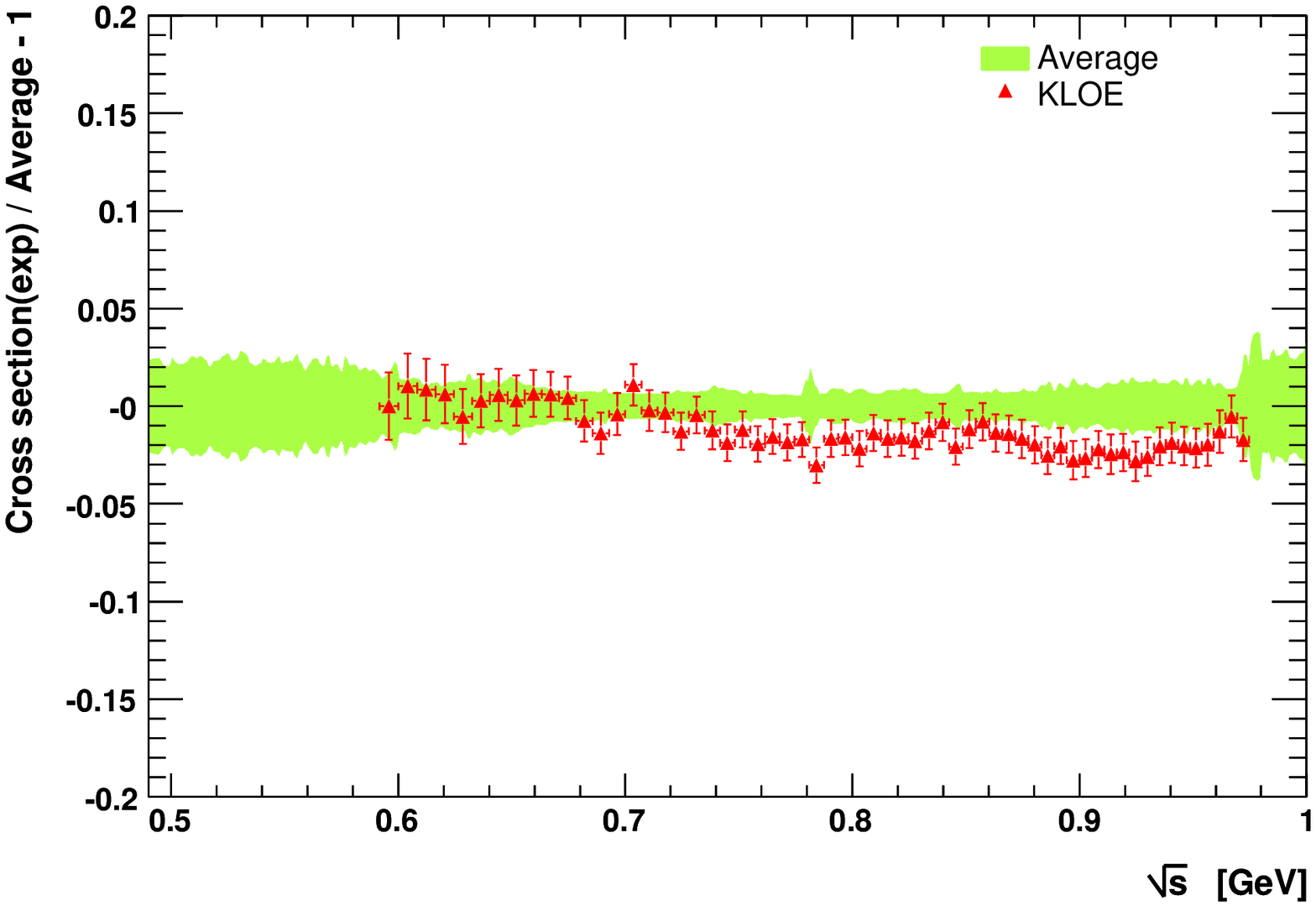}
\vspace{0.1cm}
\includegraphics[width=60mm]{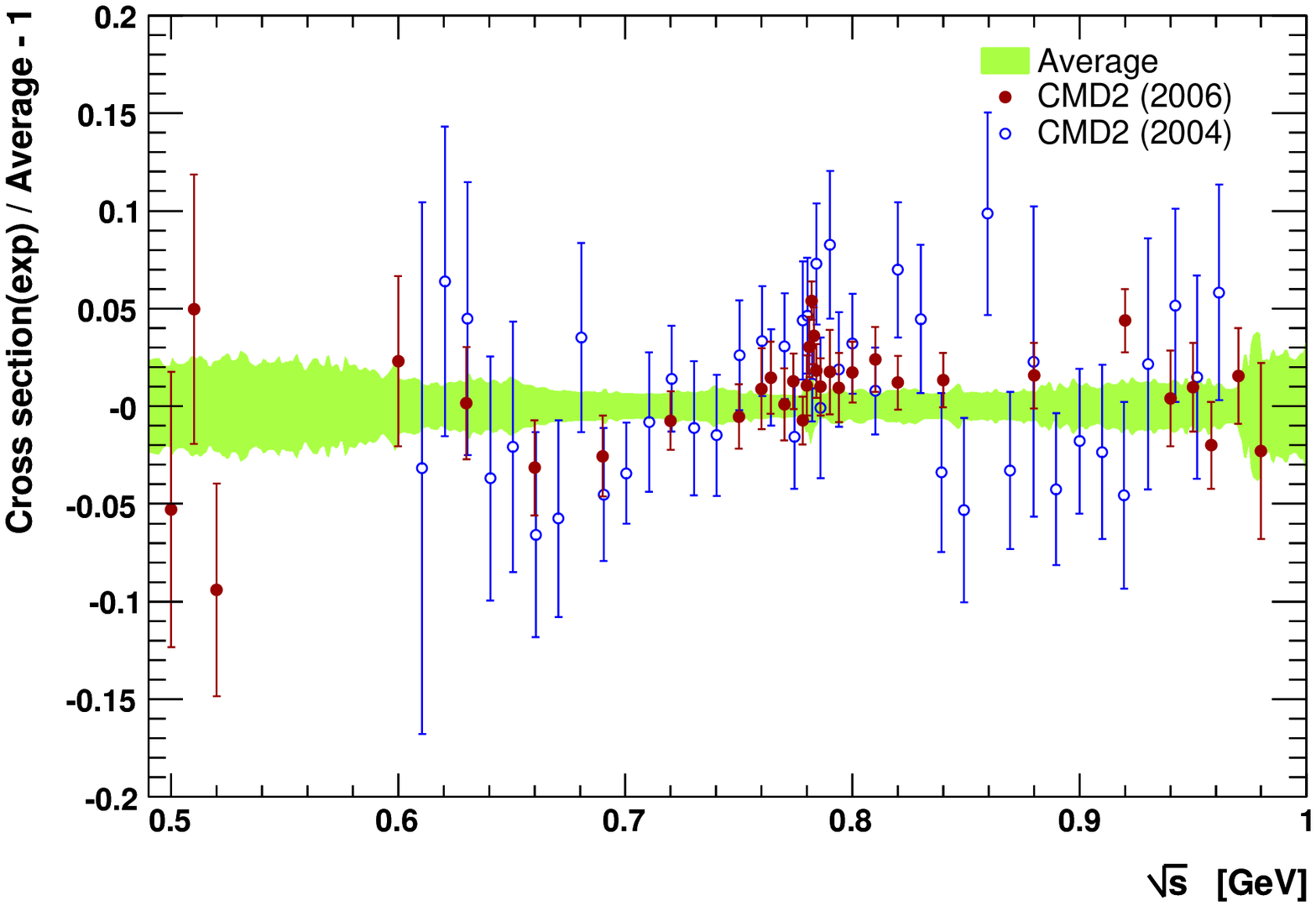}
\includegraphics[width=60mm]{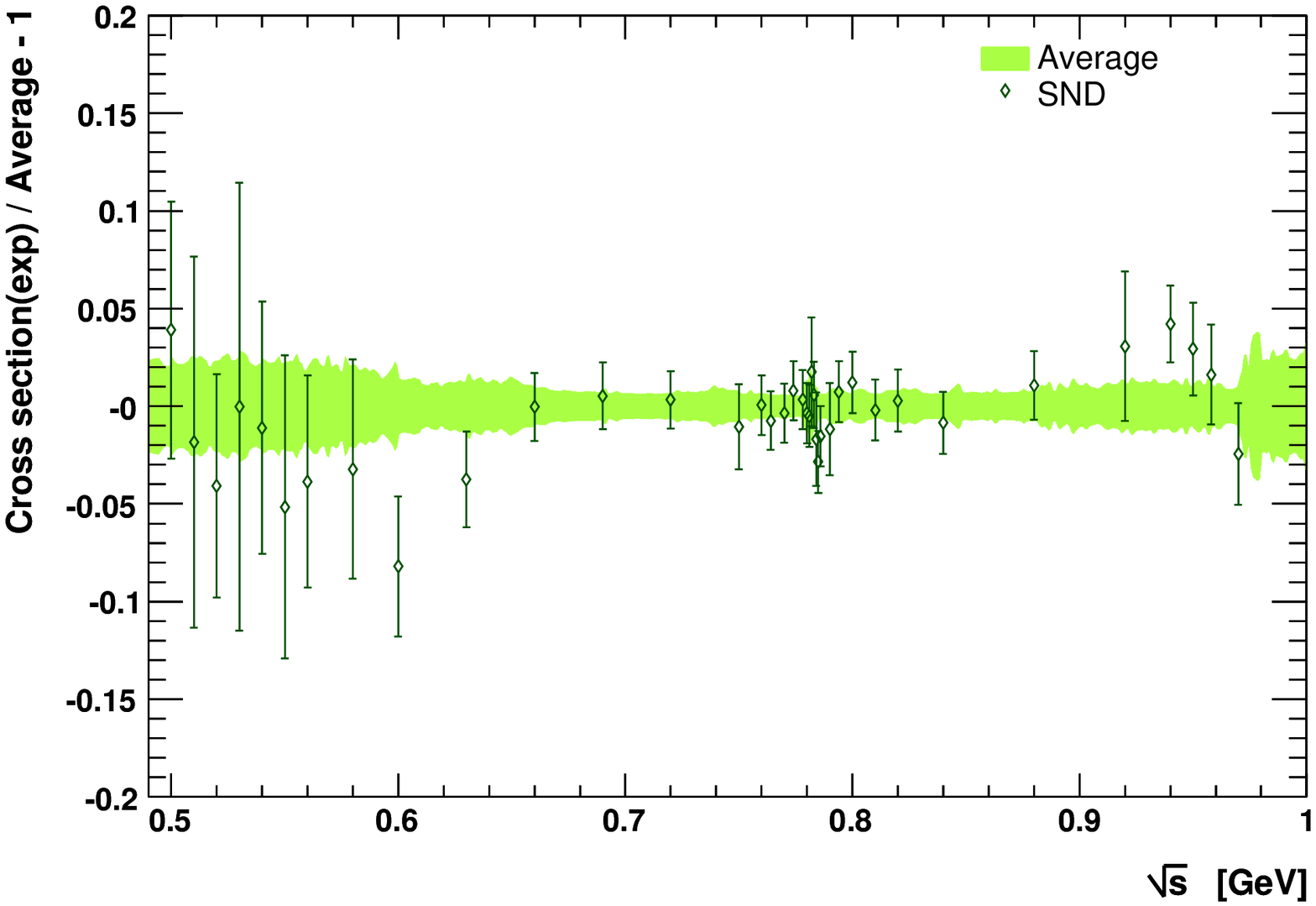}
\figcaption{\label{fig:comp} \small 
    Relative cross section comparison between individual experiments   
    (symbols) and the HVPTools average (shaded band) computed from all 
    measurements considered. Shown are \babar\ (top left), KLOE (top right), 
    CMD-2 (bottom left) and SND (bottom right).}
\end{center}

\begin{multicols}{2}

The relative differences between \babar\, KLOE, CMD-2, SND,
and their average are given in Fig.~\ref{fig:comp}. Fair agreement is observed,
though with a tendency to larger (smaller) cross sections above $\sim$$0.8\gev$
for \babar\ (KLOE). These inconsistencies (among others) lead to the error 
rescaling shown in Fig.~\ref{fig:chi2}. 

\subsection{Results for \amuhadLOpp}

A compilation of results for \amuhadLOpp\ for the various sets of experiments 
and energy regions is given in Table~\ref{tab:results}. The inclusion of 
the new \babar\ data significantly increases the central value 
of the integral, without however providing a large error reduction, because of 
the incompatibility between mainly \babar\ and KLOE, causing an increase of
the combined error. In the energy interval between 0.63 and 0.958\gev, the 
discrepancy between the \amuhadLOpp\ evaluations from KLOE and \babar\ 
amounts to $2.0\sigma$.
Since \babar\ is the only experiment covering the entire energy region between 
$2m_\pi$ and 1.8\gev, it can provide its own evaluation~\cite{babarpipi} of 
\amuhadLOpp, $514.1 \pm 2.2_{\rm stat} \pm 3.1_{\rm syst}$.~\footnote{When not
specified, the \amu\ values are given in units of $10^{-10}$.}

\section{Multihadronic Contributions}

We also reevaluate the $e^+e^-\rightarrow\pi^+\pi^-2\pi^0$ contribution to 
\amuhadLO. It is found that the CMD-2 data used previously have been superseded
 by modified or more recent, but yet unpublished data~\cite{logashenko}, 
recovering agreement with the published SND cross sections~\cite{sndpp2pi0}.

\begin{center}
  \tabcaption{\label{tab:results} \small
    Evaluated \amuhadLOpp\ contributions from the $ee$ data 
    for different energy intervals and experiments. Where two errors are given,
    the first is statistical and the second systematic. The last value in
    parentheses is the total error). Also given is the $\tau$-based result. 
}
\small
\begin{tabular}{ccc}\hline\hline
  $\sqrt{s}$ (GeV)  &  Exp.   &  \amuhadLO $[\pi\pi]$ ($10^{-10}$) \\
\hline
$2m_{\pi^\pm}-.3$ & $ee$ fit  & $0.55\pm 0.01$ \\ 
$.30-.63$   & Comb. $ee$  & $132.6 \pm 0.8 \pm 1.0$ ($1.3$) \\
$.63-.958$ & CMD2 03        & $361.8 \pm 2.4 \pm 2.1$ ($3.2$) \\
             & CMD2 06        & $360.2 \pm 1.8 \pm 2.8$ ($3.3$) \\
             & SND  06        & $360.7 \pm 1.4 \pm 4.7$ ($4.9$) \\
             & KLOE 08        & $356.8 \pm 0.4 \pm 3.1$ ($3.1$) \\
             & BABAR       & $365.2 \pm 1.9 \pm 1.9$ ($2.7$) \\
             & Comb. $ee$   & $360.8 \pm 0.9 \pm 1.8$ ($2.0$) \\
$.958-1.8$  & Comb. $ee$   &  $14.4 \pm 0.1 \pm 0.1$ ($0.2$) \\ 
\hline
Total        & Comb. $ee$   & $508.4 \pm 1.3 \pm 2.6$ $(2.9)$\\
Total        & Comb. $\tau$
                            & $515.2 \pm 2.0 \pm 2.7$ $(3.4)$ \\
\hline\hline
\end{tabular}
\end{center}
 
Since the new data are unavailable, we discard the obsolete CMD-2 data from 
the $\pi\pi2\pi^0$ average, finding 
\amuhadLOppzz $=17.6\pm0.4_{\rm stat}\pm1.7_{\rm syst}$ (compared to 
$17.0 \pm 0.4_{\rm stat} \pm 1.6_{\rm syst}$ when including the obsolete CMD-2 
data). The corresponding cross section measurements and HVPTools 
average are shown in Fig.~\ref{fig:4pi}. From the still preliminary \babar\ 
results~\cite{solodov-phipsi} it is clear that the region above 1.4 GeV is
still underestimated at the present state, as corroborated by $\tau$ 
data~\cite{dehz02}.

\begin{center}
\includegraphics[width=70mm]{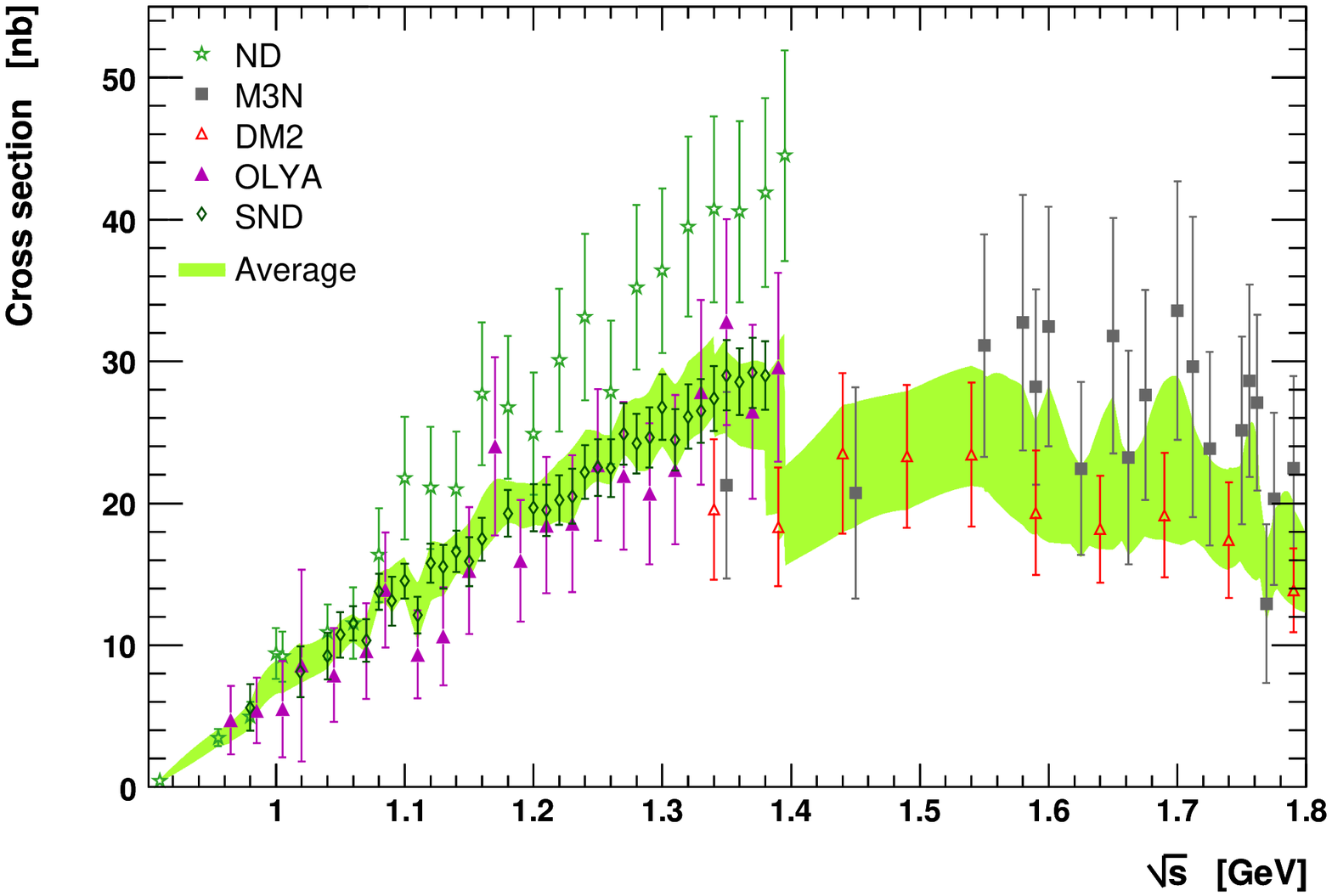}
\figcaption{\label{fig:4pi} \small 
           Cross section measurements for $e^+e^-\rightarrow\pi^+\pi^-2\pi^0$  
            used in the calculation of \amuhadLOppzz. The shaded band depicts 
            the HVPTools interpolated average within $1\sigma$ errors.
            The individual measurements are referenced in~\cite{dehz02}. }
\end{center}

\section{Results and comparison to experiment}

Adding to the $ee$-based \amuhadLOpp\ and \amuhadLOppzz\ results the remaining 
exclusive multi-hadron channels as well as perturbative QCD~\cite{md_tau06}, 
we find for the complete lowest-order hadronic term
\begin{equation}
      a_\mu^{\rm had,LO}~[ee] \:=\: 695.5 \pm 4.0_{\rm exp}\pm  0.7_{\rm QCD}~(4.1_{\rm tot})\,.
\end{equation}
It is noticeable that the error from the \pp\ channel now equals the one from 
all other contributions to \amuhadLO.

Adding further the other contributions (given in Section 2),
we obtain the Standard Model prediction (still in $10^{-10}$ units)
\begin{equation}
      a_\mu^{\rm SM}~[ee] \:=\: 11\,659\,183.4 \pm 4.1 \pm 2.6 \pm 0.2~(4.9_{\rm tot})\,,
\end{equation}
where the errors have been split into lowest and higher-order hadronic, and 
other contributions, respectively. The $a_\mu^{\rm SM}~[ee]$ value deviates 
from the experimental average~\cite{bnl}, 
$a_\mu^{\rm exp}=11\,659\,208.9 \pm 5.4 \pm 3.3$~\footnote{The $g-2$ 
measurement is 
obtained from the ratio of two frequencies and needs as input the ratio of the
muon to the proton magnetic moments. The latter ratio is derived from muonium
hyperfine splitting, and its value has been updated~\cite{mohr} after the 
E-821 publication. The new value produces a shift of $+0.92~10^{-10}$ of the
\amu value.},
by $25.5 \pm 8.0$ ($3.2\sigma$). For comparison the difference obtained with
the updated $\tau$ analysis is $15.7 \pm 8.2$ ($1.9\sigma$).

A compilation of recent SM predictions for \amu\ compared with the experimental
result is given in Fig.~\ref{fig:amures}. The \babar\ results are not yet 
contained in previous evaluations. The result by HMNT~\cite{hmnt} contains 
older KLOE data~\cite{kloe04}, which have been superseded by more recent 
results~\cite{kloe08} leading to a slightly larger value for \amuhadLO.

\begin{center}
\includegraphics[width=70mm]{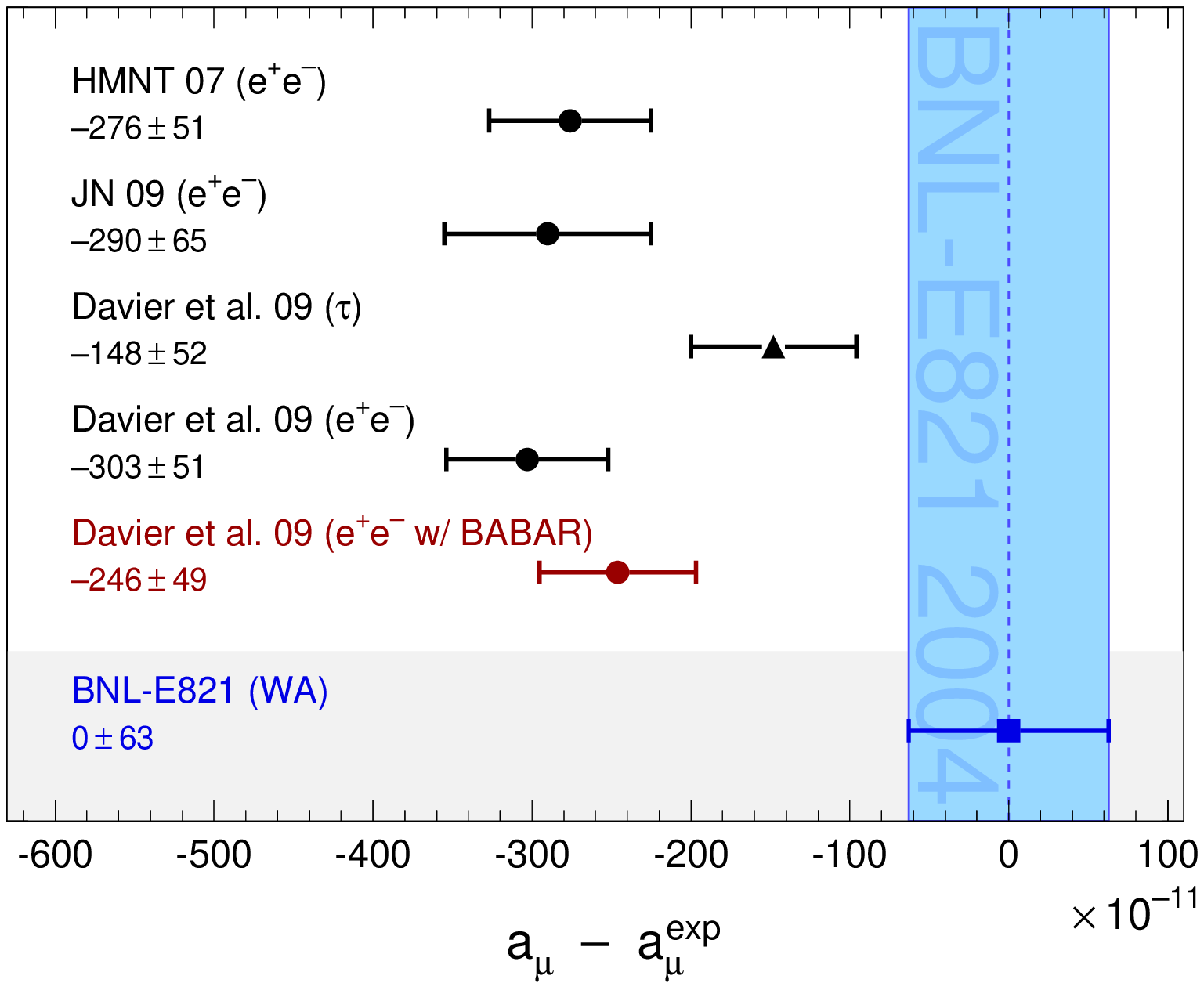}
\figcaption{\label{fig:amures} \small
   Compilation of recent results for $a_\mu^{\rm SM}$,
   subtracted by the central value of the experimental average~\cite{bnl}.
   The shaded vertical band indicates the experimental error. 
        The SM predictions are taken from: HMNT 07~\cite{hmnt}, 
        JN 09~\cite{jeger-nyff},
       Davier \ea\ 09~\cite{new-g-2-tau} ($\tau$-based and $ee$ before \babar),
        and the $ee$-based value~\cite{new-g-2-ee} including \babar.}
\end{center}

\section{Conclusions: discussion and perpectives}

The following concluding remarks can be made:

\begin{itemize}

\item
The first point to emphasize is the better consistency between the $ee$ and
$\tau$ analyses, resulting from the improved IB corrections and the \babar\
results. There is still a difference of $(6.8\pm2.9_{ee}\pm3.4_\tau)$
(1.5$\sigma$) in the $\pi\pi$ channel, but it can be 
considered as reasonable. The discrepancy was $2.9\sigma$ before, reduced
to $2.4\sigma$ after the $\tau$ update. The other major difference 
affecting the two estimates is in the $2\pi2\pi^0$ channel, 
$(3.8\pm1.7_{ee}\pm1.4_\tau)$ (1.7$\sigma$). For this case we have 
seen that a better measurement with $e^+e^-$ will likely move the $ee$ result 
closer to the $\tau$ value.

\item
The internal consistency of the $ee\rightarrow\pi\pi$ data is only fair, as
a discrepancy is observed between the \babar\ and KLOE results, which is 
increasing with energy. The difference on the $\rho$ peak, about 3\% (beyond
the respective systematic errors of 0.5\% and 1.1\%), is the most damaging for
the dispersion integral. Here CMD-2 agrees well with \babar, while SND lies
between \babar\ and KLOE. However, the accuracy of both CMD-2 and SND is not 
enough to resolve the issue.  

\item
While the \babar\ measurement is explicitly done at NLO (including one 
additional ISR or FSR photon), it is insensitive to the Monte Carlo NLO 
generation. The situation is different for KLOE which relies on Phokhara for
the ISR radiation function. Using the ISR process $ee\rightarrow\mu\mu$ \babar\
has been able to verify that Phokhara provides the right answer to an accuracy
of 1.1\%, however in a kinematic region (very hard ISR photons, 
$x=1-s/s_0>0.9$, where $s_0$ is the square of the $ee$ CM energy) far 
from that of KLOE ($0.09<x<0.66$). A measurement of the muon ISR process by
KLOE has been considered since some time and would be of considerable help to
validate their approach. 

\item
Although the accuracy of the \babar\ results is similar to that of the combined
previous $ee$ experiments, the gain in precision that could have been hoped for
was not realised because of the remaining discrepancy, essentially with KLOE.

\item
All approaches now yield a deviation from the direct measurement, however at
different levels depending on what $\pi\pi$ input data are used: $2.4\sigma$ 
with \babar\ alone, $3.2\sigma$ with all $ee$ data, $3.7\sigma$ with $ee$ not 
including \babar, $2.9\sigma$ with $ee$ not including KLOE, and $1.9\sigma$ 
with $\tau$ alone.

\item
Considering these results one can say there is some evidence for a deviation
at the $3\sigma$ level. The significance is still not enough to establish a
breakdown of the Standard Model in the muon $g-2$, {\it i.e.} a contribution 
from new physics. However it is a quantitative and very valuable information 
which will help to constrain the new physics, if it is found at the LHC.

\end{itemize}

This is the present situation. It should and will evolve as new results and
new initiatives are taking place:

\begin{itemize}

\item
The hadronic spectral functions will continue to be refined as new results 
are expected from \babar\ in the multihadronic channels, from KLOE ($\pi\pi$)
with improved approaches (preliminary results are already available for the 
large-angle ISR analysis~\cite{mueller-phipsi}), and from the upgraded CMD-3
and SND detectors at the higher-energy VEPP-2000 collider~\cite{loga-phipsi}.

\item
The theory error is still dominated by the uncertainty on the HVP contribution
($4.1$) , but it is now close to that of the hadronic LBL part 
($2.6$)~\footnote{The estimate given in Ref.~\cite{nyff-phipsi} is 
larger ($40$).}. Since the latter contribution is unlikely to be known more 
precisely in the short-term it will eventually be the theory show-stopper.

\item
But the real limitation at the moment is the $g-2$ measurement itself. The
uncertainty reached by E-821 is $6.3$, larger than the full theory error
($4.9$). It is therefore mandatory to pursue these measurements in order to 
reach higher precision. The factor 20 in precision obtained at BNL over the
pioneering measurements performed at CERN has permitted to reach the 
electroweak scale in this process. Another factor of 4, as anticipated by
the new proposal~\cite{FNAL-phipsi} submitted to Fermilab, or with the
JPARC project~\cite{JPARC-phipsi}, will definitely 
provide quantitative information as we move into the new physics territory.

\end{itemize}

\section*{Acknowledgements}

I would like to thank A.~H\"ocker, B.~Malaescu, G.~L\'opez Castro, X.H.~Mo, 
G.~Toledo S\'anchez, P.~Wang, C.Z.~Yuan, and Z.~Zhang for our fruitful 
collaboration, and C.Z.~Yuan and his colleagues for running a perfect workshop.
\\ \\

\end{multicols}

\clearpage


\begin{thebibliography}{90}
\bibitem{hvptools} The HVPTools source code and database can be made publicly 
                 available by contacting the authors of Ref.~\cite{new-g-2-ee}.
\end{thebibliography}
\end{document}